# Optically gated terahertz-field-driven switching of antiferromagnetic CuMnAs

J. J. F. Heitz[1,2], L. Nádvorník[1,2,3,*], V. Balos[1,2], Y. Behovits[1,2], A. L. Chekhov[1,2], T. S. Seifert[1,2], K. Olejník[4], Z. Kašpar[3,4], K. Geishendorf[3,4], V. Novák[4], R. P. Campion[5], M. Wolf[2], T. Jungwirth[4,5] and T. Kampfrath[1,2]

1. Department of Physics, Freie Universität Berlin, 14195 Berlin, Germany
2. Department of Physical Chemistry, Fritz Haber Institute of the Max Planck Society, 14195 Berlin, Germany
3. Faculty of Mathematics and Physics, Charles University, 121 16 Prague, Czech Republic
4. Institute of Physics, Academy of Sciences of the Czech Republic, v.v.i., 162 00 Prague, Czech Republic
5. School of Physics and Astronomy, University of Nottingham, Nottingham NG7 2RD, UK.

* E-mail: nadvornik@karlov.mff.cuni.cz

We show scalable and complete suppression of the recently reported terahertz-pulse-induced switching between different resistance states of antiferromagnetic CuMnAs thin films by ultrafast gating. The gating functionality is achieved by an optically generated transiently conductive parallel channel in the semiconducting substrate underneath the metallic layer. The photocarrier lifetime determines the time scale of the suppression. As we do not observe a direct impact of the optical pulse on the state of CuMnAs, all observed effects are primarily mediated by the substrate. The sample region of suppressed resistance switching is given by the optical spot size, thereby making our scheme potentially applicable for transient low-power masking of structured areas with feature sizes of ~100 nm and even smaller.

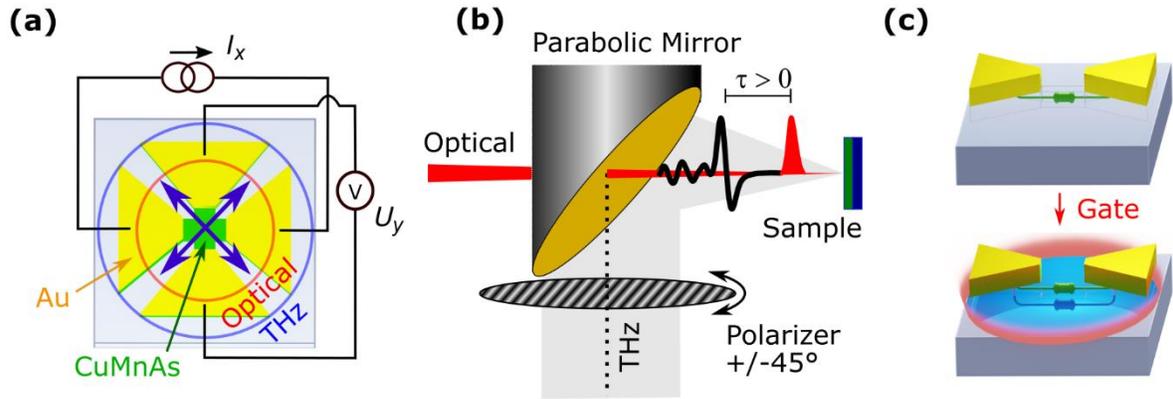

FIG. 1. Schematic of the experiment. (a) Sample structure (top view). A thin film of antiferromagnetic CuMnAs (green cross-like area) is excited by a normal incident THz pulse with circular beam cross section (blue circle) and one of two perpendicular linear polarizations (indicated by blue arrows). The structure can be gated by an additional optical femtosecond laser pulse (red circle), which excites the CuMnAs layer and the underlying substrate. To measure the electrical anisotropy of the CuMnAs layer, a cross of Au contact pads (yellow layer) allows one to apply a read current and to detect the resulting transverse voltage. The Au cross also enhances the incident THz field. (b) Schematic of the setup showing how THz radiation and optical gate pulses are focused on the sample surface. The two positions of the polarizer define the two perpendicular linear THz polarization states as shown in panel (a). (c) Principle of gating by modulating the substrate conductance. For simplicity, only one arm of the cross is shown. Without any optical excitation, the two contact pads are electrically connected through the conductance of CuMnAs (green resistor in upper sketch). Upon an excitation, the optical gate generates mobile charge carriers in the substrate which shunt the incident THz field (parallel resistor in lower sketch) and, thus, reduce the total THz field inside the CuMnAs layer.

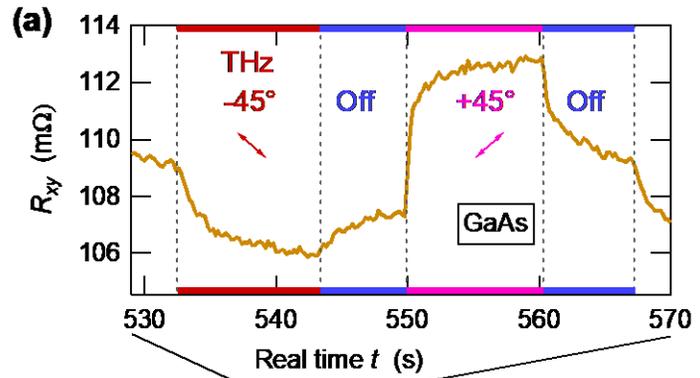

(a)

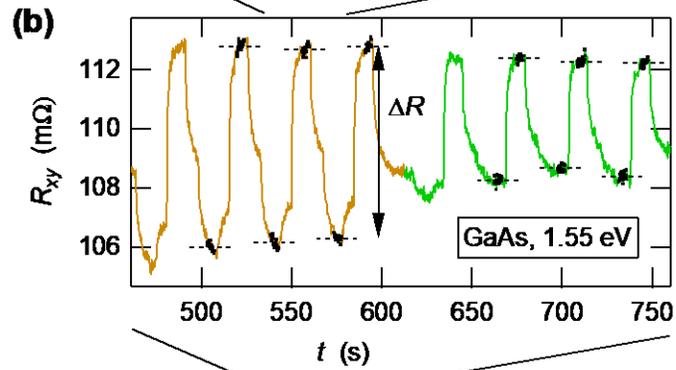

(b)

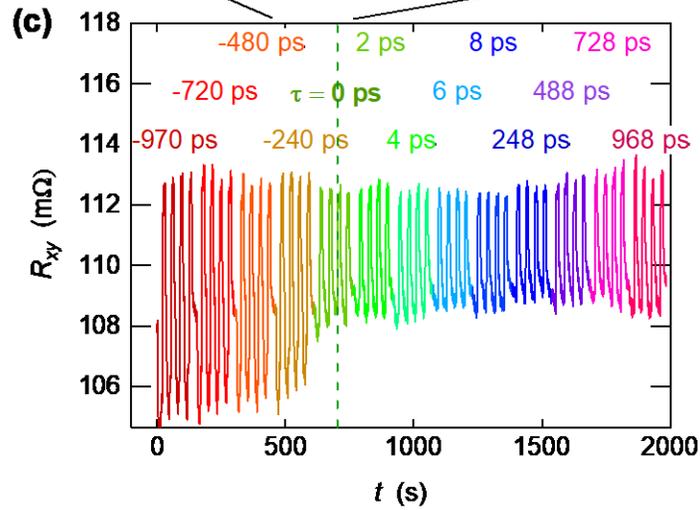

(c)

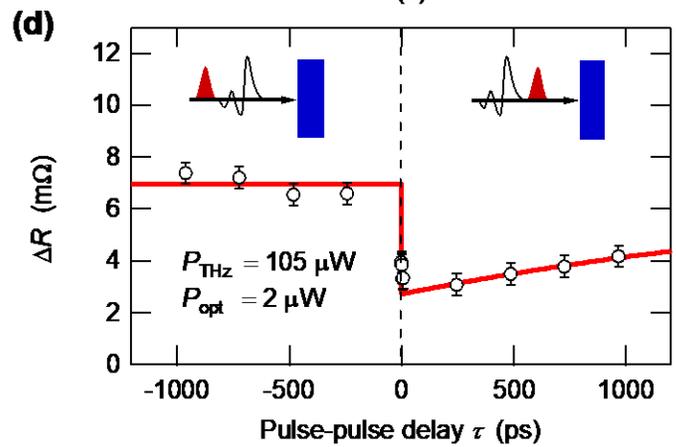

(d)

FIG. 2. Optically gated THz switching between resistivity states of a 3-µm-large CuMnAs device on a GaAs substrate. Panels (a)-(c) show the dynamics of transverse resistance vs time in increasing degree of detail. (a) Typical dynamics of the transverse resistance induced by a train of THz pulses (power $P_{THz} = 105$ µW, repetition rate 1 kHz) taken over one switching cycle (-45° THz polarization, THz off, +45° THz polarization, THz off). (b) Temporal evolution of the transverse resistance when the THz pulses are preceded by synchronized optical gate pulses (photon energy 1.55 eV, $P_{opt} = 2$ µW, repetition rate 1 kHz). The switching-induced resistance change $\Delta R$ is indicated by the double arrow. Data points highlighted in black are averaged over, and the average of each group is indicated as a black dashed line. At real time $t < 612$ s (orange data), the optical excitation occurs after the THz switching pulse (THz write-optical gate pulse delay $\tau < 0$ ps), whereas at $t > 612$ s (green data), optical excitation temporally overlaps with the THz pulse ($\tau \sim 0$ ps). (c) The same data as in panel (b), but over a larger time window. The delay $\tau$, corresponding to each color-labeled block of data, is stated above the traces. (d) Resistance modulation $\Delta R$ as inferred from panel (c) vs delay $\tau$. Each data point is the mean of $\Delta R$ of the three last switching cycles per set of four (color coded) in (b). The error bars are the standard deviation of data at $\tau < 0$. The red curve is a mono-exponential fit with a relaxation time of 2.3 ns.

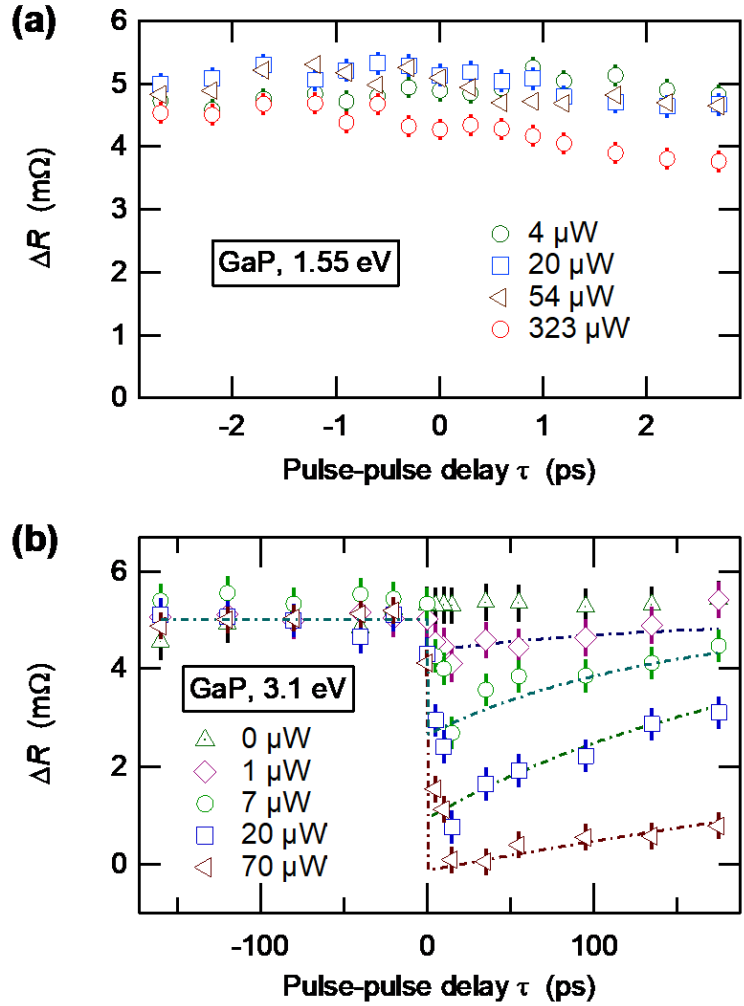

FIG. 3. Impact of the gate wavelength and substrate on the THz switching of a 2-µm-large CuMnAs device on a GaP substrate. (a) Resistance modulation $\Delta R$ vs delay $\tau$ between optical gate and THz switching pulse for different gate powers $P_{opt}$ with gate photon energy 1.55 eV (below the 2.25 eV electronic bandgap of GaP) and $P_{THz}$ = 590 µW. (b) Same as panel (a), but for a gate photon energy of 3.1 eV, above the bandgap of GaP. Note that the optical power $P_{opt}$ required for full suppression (70 µW, brown triangles) is ~5 times smaller than the maximum power applied for below-band-gap excitation of the substrate (323 µW, red circles in panel (a)). The data presented are the mean of each $\Delta R$ of the last three switching events per group as shown in Fig. 2b. The error bars are the standard deviation of data at $\tau < 0$.

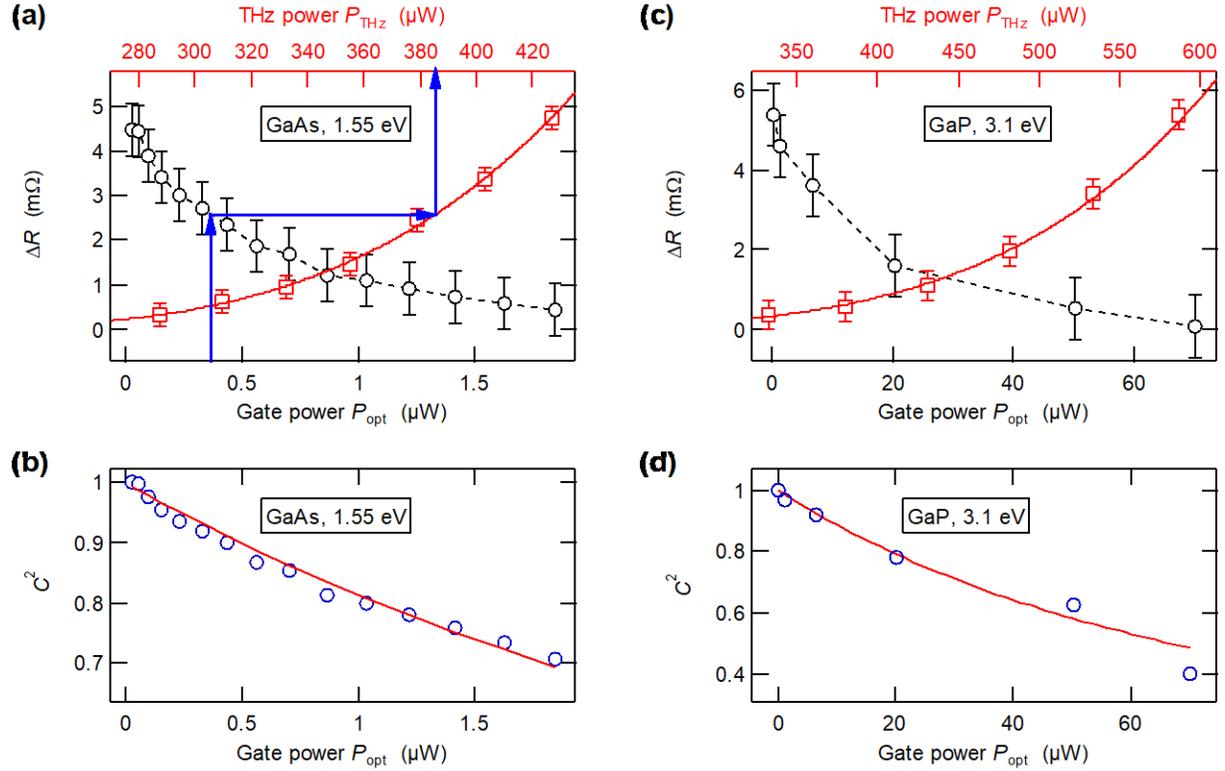

FIG. 4. Power dependence and optical gating mechanism. (a) Dependence of THz resistance modulation $\Delta R$ (see Fig. 2b) on the gate power $P_{opt}$ in a 4-µm-large CuMnAs device on GaAs substrate for the constant THz power of 426 µW and delay of $\tau = 15$ ps, that is, a preceding optical gating pulse (black symbols, bottom abscissa) and on the THz power $P_{THz}$ (red symbols, top axis, $P_{opt} = 0$). The red solid line is a fit based on $P_{THz} = a(\Delta R)^b + c$ with fit parameters $a$, $b$ and $c$. (b) Experimentally determined (blue symbols) and modeled (red solid line) factor $C^2$ that quantifies how strongly the gate pulse attenuates the THz power arriving in the CuMnAs layer as a function of the gate power $P_{opt}$. The experimental values were extracted from panel (a) (see blue arrows and text), whereas calculated values are derived from a simple parallel photoconductance model (see Fig. 1c and text). (c,d) Same as panels (a,b), but for a 2-µm-large CuMnAs device on a GaP substrate, a gate photon energy of 3.1 eV, a maximum THz power of 586 µW and a delay of $\tau = 35$ ps. The data presented are the mean of the three switching events per group as shown in Fig. 2b. The error bars are the standard deviation of data at $\tau < 0$.

| Parameter | GaAs | GaP |
|---|---|---|
| $n_S$(1 THz) (Refs. 1,2,3) | 3.6 | 3.3 |
| Gate photon energy $\epsilon_{opt}$ (eV) | 1.55 | 3.1 |
| $n_S(\epsilon_{opt})$ (Ref. 4) | 3.68+0.09i | 4.20+0.27i |
| $n_{AFM}(\epsilon_{opt})$ (Ref. 5) | 3.7+2.0i | 2.5+2.2i |
| $\mu$ (cm$^2$ V$^{-1}$ s$^{-1}$) (Refs. 6,7,8,9) | 7500 | 250 |
| $T_{opt}$ (%) | 13.8 | 2.7 |
| $G_{AFM}$ (mS) (Ref. 10 and Fig. S1) | 40 | 50 |
| $w_{opt}$ (μm) | 120 | 27 |

TABLE 1. Parameters for the parallel-photoconductor model for the two substrates used. The spot diameter $w_{opt}$ of the gate beam is a fit parameter, and transmittance $T$ is calculated as described in the main text.

# I. INTRODUCTION

Information storage using antiferromagnetic (AFM) order has potential advantages compared to ferromagnets. (i) No stray field is present, allowing for a higher integration density of memory cells.[11] (ii) Potentially fast switching of the Néel vector is possible because angular momentum transfer is not required[12,13] and zone-center magnons extend into the terahertz (THz) range.[14] Finally, (iii) robust retention of information is guaranteed because of the negligible impact of high magnetic fields up to 60 T.[15,16]

The last feature also poses a challenge because very strong magnetic fields may be required to switch the AFM order[17]. However, recent theoretical[18] and experimental works paved the way to the manipulation of the AFM order[19] by electrical currents in inversion-asymmetric AFM materials such as CuMnAs[20,21,22,23] and $Mn_2Au$[24,25]. Charge-current pulses were shown to reproducibly switch CuMnAs devices (such as in Fig. 1a) between different resistance states, which persisted for tens of seconds at room temperature and much longer at lower temperatures.[26,27] The devices can be considered multistate memory units, potentially permitting even neuromorphic operations. Interestingly, switching of CuMnAs was also achieved by free-space sub-picosecond electric-field transients driving charge currents at THz frequencies (see Fig. 1b)[10]. Recent achievements demonstrated purely optical switching between resistance states of CuMnAs without any net electric current[27].

Note that AFM CuMnAs thin-films can be grown directly on common semiconductor substrates such as Si, GaAs or GaP[21,27,28], which enable new functionalities. An important example is the variation of the shunt resistance of the substrate, which may modulate the switching process in CuMnAs (see Fig. 1c). A reduction of the substrate resistance can be achieved straightforwardly by excitation with an optical femtosecond gate pulse that generates quasi-free electrons and holes. Such transient conductivities were already successfully exploited to modulate the response of THz metamaterials[29]. In our setup, the reduced substrate resistance shorts the subsequently arriving THz electric field and, thus, should reduce the current driven in the AFM layer (see Fig. 1c), without the need for additional microstructuring. It remains to be shown whether this effect is sufficient to suppress the switching process and whether the direct impact of the optical pulse on the CuMnAs layer is negligible.

In this work, we study the interplay between THz-driven switching of a CuMnAs thin-film and the ultrafast optical excitation of the semiconducting substrate. For optically absorbing substrates, we find that the optical pump with powers smaller than the optical writing threshold can efficiently modulate and even completely suppress the switching process driven by the subsequently incident THz field but do not destroy information already written in the CuMnAs. Our observation is consistent with a model based on screening of the THz field by the optically induced free charge carriers. Our results demonstrate that integrating AFM devices with semiconductors enables new functionalities with technological relevance.

# II. SAMPLES AND SETUP

A schematic of our experiment is shown in Fig. 1b. An AFM CuMnAs thin film (Fig. 1a) grown on top of a semiconducting substrate is excited by a pair of incident pulses at a repetition rate of 1 kHz: first the optical gate pulse and, after a delay time $\tau$, the THz switching pulse with polarization directions at 45° or -45° with respect to the vertical axis. The two THz polarizations allow one to switch the AFM device between defined resistance states[10]. The impact of the excitation is probed by the continuous measurement of the transverse sample resistance via Au contact pads (Fig. 1a).

Two kinds of samples are studied: thin-films of AFM CuMnAs (thickness 50 nm, sheet conductance 40 mS on GaAs and 50 mS on GaP) on a substrate (thickness 500 μm) of GaAs (optical bandgap 1.43 eV at room temperature) and GaP (2.25 eV). These substrate materials allow us to control the amount of energy that is absorbed in the substrate from the optical gate pulse by varying the gate wavelength.

As indicated by Fig. 1a, the CuMnAs layer was patterned and partly etched out to form a cross structure (side length of the square central region 2-4 µm). The Au contact pads (length 330 µm, width 210 µm), which consist of an Au layer (thickness 80 nm) deposited on a Cr wetting layer (5 nm), are tapered toward the cross in the last third of the length and serve two purposes: First, they act as an antenna that enhances the incident THz electric field in the cross (blue arrows in Fig. 1a) by a factor of typically ~20.[10] Second, they serve as contacts to detect the electrical anisotropy of the sample (Fig. 1a), since we apply a DC read current $I_x$ (1…2 mA) along the horizontal bar of the cross pattern and measure the transverse voltage $U_y$ at a read-out rate of 5 Hz. Variations $\Delta R_{xy}$ of the transverse resistance $R_{xy} = U_y/I_x$ quantify changes in the electrical anisotropy of the sample[10] induced by the THz switching and optical gate pulse.

Intense THz pulses (Fig. 1b) are generated by optical rectification of ultrashort optical laser pulses (800 nm, duration stretched to 80 fs, energy 6 mJ, repetition rate 1 kHz) in a LiNbO$_3$ crystal, providing THz pulses with peak electric fields up to 1 MV/cm and durations of ~1 ps.[30] The polarization state of the THz beam is controlled by a wire-grid polarizer ($\pm 45°$) and combined with optical gate pulses from the same laser by a parabolic mirror with a hole in its center. Both beams are focused onto the sample, forming spots with diameters of <1000 µm (THz) and roughly 80 µm (optical) characterized by the pinhole method (see Supplementary Note 1). The fluences are controlled by a combination of optical waveplates with polarizers and neutral-density filters (optical) and the optical pump power incident on the LiNbO$_3$ crystal (THz). An example of a typical THz waveform (spectrum 0.3-2.5 THz, peak at 1 THz) and a schematic of an optical pulse are shown in Fig. 1b. While the duration of the THz electromagnetic transient is smaller than 1 ps, the duration of the optical gate pulse is approximately 40 fs.

The sample structure is excited by a train of pulse pairs at a repetition rate of 1 kHz. As indicated by Fig. 1b, each pair consists of a THz switching pulse (polarization directions at $\pm 45°$) and an optical gate pulse (repetition rate of 1 kHz). The delay $\tau$ between optical and THz pulse can be set between -1 ns and +1 ns by a mechanical translation stage. Here, a positive/negative $\tau$ means that the THz pulse arrives after/before the optical pulse at the sample, respectively (see Fig. 2d). We emphasize that the optical fluences used here are at least an order of magnitude weaker than the fluence that would be needed to switch the AFM directly[27]. The experiments are conducted at room temperature under ambient conditions. More details of the optical and electrical setup can be found in Supplementary Note 1.

### III. RESULTS

### A. Typical data

Figure 2 depicts a typical example of the measured transverse resistance $R_{xy}$ of CuMnAs on a GaAs substrate vs real time. The width of the shown time windows ranges from 40 s (Fig. 2a) over 300 s (Fig. 2b) to 2000 s (Fig. 2c). The power of the THz and optical pulse trains was set to $P_{\text{THz}} = 105$ µW and $P_{\text{opt}} = 2$ µW, respectively.

Figure 2a displays one switching cycle where the gate pulse arrives after the THz switching pulse ($\tau = -240$ ps). The course of the transverse resistance signal $R_{xy}$ has the characteristic shape of a switching cycle as observed in previous works[10,20,27,31]. Exposure with -45°-polarized THz pulses causes switching, while the absence of the THz pulses ("THz off") results in a relaxation of the signal towards a baseline. Exposure with +45°-polarized THz pulses switches the sample resistance state into the opposite direction away from the baseline, to which the signal relaxes back again after removing the THz excitation.

Figure 2b shows an extended time window that contains two signal groups. Each of these groups contains 4 switching cycles and is marked by a different color. The first four-group (brown) was taken for $\tau = -240$ ps (gate arrives after THz pulse), whereas the second four-group (green) was measured for coinciding arrival times of THz and gate pulses ($\tau = 0$ ps). We observe a distinct reduction of the resistance modulation by a factor of about 2. The resistance modulation $\Delta R$ is the maximum variation of the transverse resistance $R_{xy}$ during the corresponding full switching cycle (see arrow in Fig. 2b).

Figure 2c displays an extended window of measurement time and, accordingly, a wider range of THz-gate delays $\tau$ than in Fig. 2b. For negative delays $\tau < 0$ (gate after THz pulse) there are no significant variations of the resistance modulation. At $\tau = 0$ ps, a reduction of resistance modulation is observed, which slightly recovers over the subsequent 1000 ps of optical-THz delay. To analyze this behavior, we extract the resistance modulation from the last 3 cycles of each 4-group of Fig. 2c. Note that the first cycle is omitted from this analysis as it is affected by the baseline shift from the previous parameter set. The average of these three contrast values for each four-group yields one data point. As an estimate of the error bar, we take the standard deviation of data for $\tau < 0$.

The resulting $\Delta R$ vs pump-probe delay is shown in Fig. 2d. The gate-induced reduction of the resistance modulation is significantly bigger than the error bar. The gate induces a step-like decrease of the resistance modulation when the optical pulse starts preceding the THz pulse with subsequent relaxation on a nanosecond time scale. The underlying dynamics can be well described by a step function times a mono-exponential decay with a time constant of 2.3 ns and a constant background (see red curve in Fig. 2d).

### B. Impact of the gate pulse

Figures 2c,d demonstrate that the optical gate is capable of suppressing the contrast of switching over one THz-driven switching cycle. Note that the optical gate (photon energy of 1.55 eV) used in the measurements, shown in Fig. 2, can excite both the GaAs substrate (electronic band gap of 1.43 eV) and the CuMnAs film.

To address the role of the substrate, we change the substrate to GaP (bandgap of 2.25 eV) and repeat the experiment with a gate photon energy below (1.55 eV, Fig. 3a) and above (3.1 eV, Fig. 3b) the GaP bandgap. For below-gap gate excitation (Fig. 3a), no significant change in the resistance modulation is observed relative to the error bars of the measurement (Fig. 2a). Thus, the direct excitation of CuMnAs has a minor impact on the switching dynamics, which most likely also holds true for the samples grown on GaAs.

However, for above-gap gate excitation (Fig. 3b), a distinctly different behavior of the resistance modulation vs THz-optical delay is found. We observe a strong reduction of the resistance modulation which increases with gate power. At the maximum gate power of $P_{\text{opt}} = 70$ µW, nearly complete suppression is achieved for delays below ~20 ps. We conclude that suppression of the switching signal is related to resonant electronic excitation of the substrate of the CuMnAs device, while direct excitation of the CuMnAs and Au films (both at 1.55 eV and 3.1 eV) plays a minor role. The time scales on which the resistance modulation recovers are discussed further below.

### C. Power dependence

To characterize how strongly the optically excited substrate modifies the THz excitation process, we measure the resistance modulation $\Delta R$ as a function of the THz power $P_{\text{THz}}$ and the gate power $P_{\text{opt}}$ (Fig. 4a). Without the optical gate pulses, we observe that the resistance modulation $\Delta R$ increases monotonically and strongly superlinear with increasing THz power (red curve in Fig. 4a), consistent with the observations in Ref. 10.

In a second measurement, the THz power is held at the maximum available value of $P_{\text{THz}}^{\max} = 426$ μW, and the gate-pulse train is switched on with a delay of $\tau = 15$ ps. We see that the resistance modulation $\Delta R$ decreases monotonically when the gate power is increased (black curve in Fig. 4a). At a gate power above 1.8 μW, $\Delta R$ is suppressed by more than 90%. An analogous behavior of $\Delta R$ vs $P_{\text{THz}}$ and $P_{\text{opt}}$ is found for the CuMnAs device on the GaP substrate for 3.1 eV gate excitation (Fig. 4c), yet for significantly larger gate powers. This is a consequence of a combination of material parameters (Table 1) as explained by the model presented in the next section. Note that these observations demonstrate a high level of control over the resistance modulation by the gate pulse.

### D. "Parallel photoconductor" model

A mechanism through which the optically excited substrate can modify the THz excitation process is the "parallel photoconductor" model, which is schematically shown in Fig. 1c. In a qualitative picture, without the optical gate, the incident THz pulse drives a current exclusively through the AFM CuMnAs layer of conductance $G_{\text{AFM}}$ (see Fig. 1c top). However, if the THz pulse is preceded by an optical gate, the gate generates quasi-free electrons and holes in the substrate, resulting in a transient electrically conducting layer with sheet conductance $\Delta G_S(P_{\text{opt}}, \tau)$ (see Fig. 1c bottom). The transient layer shunts the CuMnAs and, thus, reduces the current through the CuMnAs layer by a factor $C$.

We extract the dependence of $C$ on the gate power $P_{\text{opt}}$ from the data of Fig. 4a: For a given $P_{\text{opt}}$ and maximum THz power $P_{\text{THz}}^{\max} = 426$ μW, we determine the equivalent THz power $P_{\text{THz}}^{\text{eq}}$ that is required to induce the same resistance modulation $\Delta R$ in the absence of the gate beam, i.e., $P_{\text{opt}} = 0$. This procedure is illustrated by the 3 blue arrows in Fig. 4a. It requires a continuous curve $\Delta R$ vs $P_{\text{THz}}$, which is obtained from a fit based on $P_{\text{THz}} = a(\Delta R)^b + c$ with fit parameters $a, b, c$. As the THz power is proportional to the integrated squared THz field, the field reduction factor $C$ is finally determined by the relationship $C^2 = P_{\text{THz}}^{\text{eq}}/P_{\text{THz}}^{\max}$. The resulting $C^2$ is plotted vs $P_{\text{opt}}$ in Fig. 4b. Starting from unity, $C^2$ decreases monotonically with increasing $P_{\text{opt}}$ from 0 to 1.8 μW.

To test our "parallel photoconductor" model (Fig. 1c), we calculate $C$ using a simple model, in which we consider the device as an infinitely extended CuMnAs thin film between two half-spaces of air (A) and the substrate (S). As the penetration depth of the gate pulse is 750 nm for GaAs at 1.55 eV and 116 nm for GaP at 3.1 eV, the thickness of the transiently conducting S layer and the CuMnAs film (50 nm) is smaller than the skin depth (>1 μm) of the THz field in the CuMnAs and substrate (see Supplementary Note 2). Therefore, the THz electric field is approximately constant across the thickness of the CuMnAs and the photoexcited substrate layer, allowing us to apply the Tinkham formula,[32,33] which relates the incident THz electric field $E_{\text{inc}}$ to the electric field $E_{\text{AFM}}$ inside the CuMnAs layer with the transmission coefficient

$$t(\Delta G_S) = \frac{E_{\text{AFM}}(\Delta G_S)}{E_{\text{inc}}} = \frac{2n_A}{n_A + n_S + Z_0 G_{\text{AFM}} + Z_0 \Delta G_S}. \quad (1)$$

Here, $n_A$ and $n_S$ are the refractive indices of air and unexcited substrate, respectively, and $Z_0 \approx 377$ Ω is the free-space impedance. Equation (1) shows that we can control the amplitude of the THz field $E_{\text{AFM}} = tE_{\text{inc}}$ inside CuMnAs by tuning the amplitude of $E_{\text{inc}}$ or by modifying the transient conductance

$\Delta G_S(P_{\text{opt}}, \tau)$ of the photoexcited substrate layer through variation of the energy and time delay of the optical gate pulse. It follows that $\Delta G_S$ decreases the field inside the CuMnAs by a factor of

$$C = \frac{E_{\text{AFM}}(\Delta G_S)}{E_{\text{AFM}}(0)} = \frac{t(\Delta G_S)}{t(0)} \approx \frac{1}{1 + \Delta G_S/G_{\text{AFM}}}. \quad (2)$$

The last approximation is justified because $Z_0 G_{\text{AFM}} \approx 15$ is significantly larger than $n_A + n_S \approx 4.6$ and approximately constant at 0-2 THz (see Supplementary Fig. S1 and Note 2). Photoinduced changes $\Delta G_{\text{AFM}}$ in the conductance of the CuMnAs film are neglected because they do not lead to a modification of the THz-induced resistance modulation for below-gap excitation (Fig. 3a). Directly after arrival of the pump pulses, when charge-carrier recombination is not yet relevant, the photoinduced conductance $\Delta G_S$ can be estimated using the relationship

$$\Delta G_S = e\mu \cdot \left(\frac{PT}{f\epsilon \cdot \pi w^2/4}\right)_{\text{opt}}, \quad (3)$$

where $e$ is the elementary charge, and $\mu$ is the mobility of a photogenerated electron-hole pair in GaAs or GaP. In Eq. (3), the term in the brackets with subscript "opt" equals the number of photons per optical gate pulse and per sample area that get absorbed by the substrate layer. Accordingly, $P_{\text{opt}}$, $w_{\text{opt}}$, $\epsilon_{\text{opt}}$ and $f_{\text{opt}} = 1$ kHz denote, respectively, the mean power, beam diameter, photon energy and pulse repetition rate of the gate beam. Therefore, $\Delta G_S$ is directly proportional to $P_{\text{opt}}$. The gate-intensity transmittance $T_{\text{opt}}$ from air through the CuMnAs film to the substrate layer is calculated by the Airy formula that takes all reflection echoes in the CuMnAs into account, using the optical refractive indices of substrate, air and CuMnAs[5] (see Supplementary Note 2). The relevant parameter values of the GaAs and GaP substrates are summarized in Table 1.

We fit Eqs. (2) and (3) to the experimentally determined $C^2$ vs $P_{\text{opt}}$ by using the gate-beam diameter $w_{\text{opt}}$ at the sample surface as the only free parameter. As seen in Fig. 4b and d, good agreement between model and experiment is obtained for $w_{\text{opt}} = 120$ μm and 27 μm for GaAs and GaP substrate, respectively. These values compare favorably to the experimentally estimated spot diameter using a pinhole technique, which amounts to ~80 μm for GaAs and $\epsilon_{\text{opt}} = 1.55$ eV. For GaP and $\epsilon_{\text{opt}} = 3.1$ eV, the gate wavelength is a factor of 2 smaller, thereby suggesting a gate focus diameter that is also reduced by half to a value of ~40 μm. These values are a lower limit to the gate spot size at the sample surface because of uncertainties in the sample position. The values obtained from the model are consistent within a factor of 2 with these estimates.

This agreement supports our interpretation, suggesting that the gate-induced reduction of the resistance modulation of our samples arises from shunting by the photoconductance of optically generated charge carriers in the substrate (Fig. 1c). Our model can in particular explain the much smaller gate power that is required to reduce the THz-resistance modulation of the sample with the GaAs (Fig. 4b) vs the GaP (Fig. 4d) substrate. The difference arises from the electron mobility, which is a factor of ~30 larger for GaAs than for GaP (see Table 1).

## IV. DISCUSSION

Our interpretation of the gate-induced suppression of the resistance modulation (Fig. 1c) implies that the temporal decay of $\Delta R$ is due to the relaxation of the quasi-free electrons and holes generated by the optical gate in the substrate. Indeed, the relaxation times of $\Delta R$ seen in our experiments are compatible with time scales found in previous works: For GaAs-based samples, we find a mono-exponential decay with time constant of 2.3 ns, which agrees well with the typical range of carrier lifetimes in GaAs from several hundreds of picoseconds to nanoseconds, depending on density of in-gap states and compensation doping[34,35,36]. For GaP, the time constants of the decay of $\Delta R$ vary from 140 ps to 800 ps, depending on excitation density, and are of the same order of magnitude as time scales found for other large-gap semiconductors such as 6H-SiC (ref. 37) and ZnTe[38]. We emphasize that comparison of carrier lifetimes is useful only in a qualitative sense. First, they depend sensitively on sample properties such as defects, carrier density and surface quality. Second, the observable $\Delta R$ considered here (Fig. 3) depends highly nonlinearly on the carrier density (see Fig. 4a,c) and so may not faithfully reflect the speed of the carrier decay.

For below-bandgap excitation (GaP substrate, $\epsilon_{\text{opt}} = 1.55$ eV), no impact of the optical gate pulse on the THz switching process is observed (see Fig. 3a, $\tau > 0$). Similarly, the excitation does not modify the already written resistance state (see Figs. 2d and 3, $\tau < 0$). We can conclude that the gate-pulse energy densities used here do not reach the threshold that is needed for optical writing and erasing of resistance states in CuMnAs. Observation of such changes would require a transient increase of the CuMnAs temperature close to the Néel point[27]. In fact, we estimate the maximal energy density deposited by one optical pulse in the CuMnAs film in our experiment to be roughly 50 pJ/μm$^3$, which is approximately one order of magnitude smaller than the densities used for optical writing and erasing in Ref. 27. By comparison with these switching experiments, we estimate that the transient temperature increase of CuMnAs reaches a few 10 K. Therefore, we consider the optical gate-pulse energies used here to be optimal: We control the THz field strength in the CuMnAs film without directly changing the resistance state of the CuMnAs film.

The time scale over which the AFM memory stays protected against the THz switching is determined by the lifetime of the photo-induced carriers. The latter can be set over a wide range from sub-picoseconds to nanoseconds by standard semiconductor methods, for example control of the impurity concentration and growth temperature of the substrate[39,40].

In a wider application perspective, our results demonstrate the integration of an AFM memory with a powerful semiconductor functionality. It suggests a pathway to achieve THz-field-induced resistance switching of selected regions of the CuMnAs film, an application which is analogous to an "optical-gate transistor". Illumination with a structured gate-beam cross-section protects the illuminated regions from being switched by the global THz field without affecting the already stored information. For free-space gate beams, the smallest feature size and, thus, footprint of a bit is given by half the gate wavelength, that is, $\lambda_{\text{opt}}/2 = 200$ nm. This value is three orders of magnitude smaller than the THz beam diameter. Adequate and rapid structuring of light beams can be achieved with binary spatial light modulators.[41]

The integration of AFM memory functionality with logical features of semiconductors shown here and the compatibility of CuMnAs with various semiconductor substrates open the door toward a rich variety of applications. For instance, an optical excitation of memory bits, similar to our optical protection from THz writing, is already successfully employed in heat-assisted magnetic recording (HAMR) technology. Here, a magnetically hard ferromagnetic recording surface is exposed to a magnetic field of relatively large spatial extent, but the data is written only in a small region softened by thermal heating from an optical beam. Small feature sizes of the order of $50 \times 50$ nm$^2$ are routinely possible with optical nano-antennas[42]. It is interesting to compare the optical energy required to soften one magnetic bit in HAMR to the energy that would be required to excite the semiconducting substrate and protect one bit from THz-field-induced writing. Assuming a bit size of $50 \times 50$ nm$^2$, we estimate that HAMR requires an optical energy of 30 pJ/bit[43,44]. This value is more than five orders of magnitude larger than the 0.5 fJ/bit needed for bit protection in our samples (see Supplementary Note 3). Therefore, the scheme suggested in this work (Fig. 1c) is comparatively energy-efficient and takes the advantages of antiferromagnetic data storage to semiconductor electronics.

## V. CONCLUSION

Our results demonstrate an efficient control of the THz switching of a CuMnAs AFM layer by functionalizing the semiconducting substrate. When excited with an optical pulse, the substrate becomes conducting and shorts the writing current in the AFM layer. A parallel-photoconductor model can quantitatively explain our findings. Notably, we do not observe a direct influence of the optical gate absorption by the AFM layer itself. For technological applications, the lifetime of the photoinduced changes in the conductivity of the semiconducting substrate can be tuned by doping and defect density. The shown integration of AFM memory functionality with logical features of semiconductors and the compatibility of CuMnAs with various semiconductor substrates open the door toward a rich variety of applications known from dilute magnetic semiconductors [45,46] and multiferroic spintronics[47].

## ACKNOWLEDGMENTS

We thank the European Union for support through the projects CoG TERAMAG/Grant No. 681917 and ASPIN/Grant No. 766566, the Ministry of Education of the Czech Republic for Grants LM2018110 and LNSM-LNSpin and the German Research Foundation for funding through the collaborative research center SFB TRR 227 "Ultrafast spin dynamics" (project A05). J.J.F. Heitz acknowledges support by the IMPRS for Elementary Processes in Physical Chemistry

# REFERENCES

[1] J. Hebling, A. G. Stepanov, G. Almási, B. Bartal, and J. Kuhl, Applied Physics B: Lasers and Optics 78, 593 (2004).

[2] D. Grischkowsky, S. Keiding, M. van Exter, and Ch. Fattinger, J. Opt. Soc. Am. B 7, 2006 (1990).

[3] K. L. Vodopyanov and Yu. H. Avetisyan, Opt. Lett. 33, 2314 (2008).

[4] D. E. Aspnes and A. A. Studna, Phys. Rev. B **27**, 985 (1983).

[5] M. Veis, J. Minár, G. Steciuk, L. Palatinus, C. Rinaldi, M. Cantoni, D. Kriegner, K. K. Tikuišis, J. Hamrle, M. Zahradník, R. Antoš, J. Železný, L. Šmejkal, X. Marti, P. Wadley, R. P. Campion, C. Frontera, K. Uhlířová, T. Duchoň, P. Kužel, V. Novák, T. Jungwirth, and K. Výborný, Phys. Rev. B 97, 125109 (2018).

[6] J. S. Blakemore, Journal of Applied Physics 53, R123 (1982).

[7] G. E. Stillman, C. M. Wolfe, and J. O. Dimmock, Journal of Physics and Chemistry of Solids 31, 1199 (1970).

[8] H. C. Casey, F. Ermanis, and K. B. Wolfstirn, Journal of Applied Physics 40, 2945 (1969).

[9] Y. C. Kao and O. Eknoyan, Journal of Applied Physics 54, 2468 (1983).

[10] K. Olejník, T. Seifert, Z. Kašpar, V. Novák, P. Wadley, R. P. Campion, M. Baumgartner, P. Gambardella, P. Němec, J. Wunderlich, J. Sinova, P. Kužel, M. Müller, T. Kampfrath, and T. Jungwirth, Sci. Adv. 4, eaar3566 (2018).

[11] Jungfleisch, M. B., Zhang, W. & Hoffmann, A. Perspectives of antiferromagnetic spintronics. *Physics Letters A* **382**, 865–871 (2018).

[12] S. F. Maehrlein, I. Radu, P. Maldonado, A. Paarmann, M. Gensch, A. M. Kalashnikova, R. V. Pisarev, M. Wolf, P. M. Oppeneer, J. Barker and T. Kampfrath, Dissecting spin-phonon equilibration in ferrimagnetic insulators by ultrafast lattice excitation, *Science Advances,* **2018***, 4*, eaar5164

[13] K. Jhuria, J. Hohlfeld, A. Pattabi, E. Martin, A. Y. A. Córdova, X. Shi, R. L. Conte, S. Petit-Watelot, J. C. Rojas-Sanchez, G. Malinowski, S. Mangin, A. Lemaître, M. Hehn, J. Bokor, R. B. Wilson, J. Gorchon, Spin–orbit torque switching of a ferromagnet with picosecond electrical pulses. *Nat Electron* **3,** 680–686 (2020). https://doi.org/10.1038/s41928-020-00488-3

[14] T. Kampfrath, A. Sell, G. Klatt, A. Pashkin, S. Mährlein, T. Dekorsy, M. Wolf, M. Fiebig, A. Leitenstorfer, R. Huber, Coherent terahertz control of antiferromagnetic spin waves. *Nature Photon* **5,** 31–34 (2011). https://doi.org/10.1038/nphoton.2010.259

[15] Song, C. *et al.* How to manipulate magnetic states of antiferromagnets. *Nanotechnology* **29**, 112001 (2018).

[16] Kašpar, Z., Surýnek, M., Zubáč, J. *et al.* Quenching of an antiferromagnet into high resistivity states using electrical or ultrashort optical pulses. *Nat Electron* **4,** 30–37 (2021). https://doi.org/10.1038/s41928-020-00506-4

[17] M. Wang, C. Andrews, S. Reimers, O. J. Amin, P. Wadley, R. P. Campion, S. F. Poole, J. Felton, K. W. Edmonds, B. L. Gallagher, A. W. Rushforth, O. Makarovsky, K. Gas, M. Sawicki, D. Kriegner, J. Zubáč, K. Olejník, V. Novák, T. Jungwirth, M. Shahrokhvand, U. Zeitler, S. S. Dhesi, and F. Maccherozzi, Spin flop and crystalline anisotropic magnetoresistance in CuMnAs, Phys. Rev. B 101, 094429 (2020)

[18] Železný, J. *et al.* Relativistic Néel-Order Fields Induced by Electrical Current in Antiferromagnets. *Phys. Rev. Lett.* **113**, 157201 (2014).

[19] S. Schlauderer, C. Lange, S. Baierl, T. Ebnet, C. P. Schmid, D. C. Valovcin, A. K. Zvezdin, A. V. Kimel, R. V. Mikhaylovskiy, R. Huber, Temporal and spectral fingerprints of ultrafast all-coherent spin switching. *Nature* **569,** 383–387 (2019). https://doi.org/10.1038/s41586-019-1174-7

[20] Wadley, P. *et al.* Electrical switching of an antiferromagnet. *Science* **351**, 587–590 (2016).

[21] Olejník, K. *et al.* Antiferromagnetic CuMnAs multi-level memory cell with microelectronic compatibility. *Nat Commun* **8**, 15434 (2017).

[22] Krizek *et al*., Atomically sharp domain walls in an antiferromagnet, ArXiv:2012.00894 (2020).

[23] M. S. Wörnle, P. Welter, Z. Kašpar, K. Olejník, V. Novák, R. P. Campion, P. Wadley, T. Jungwirth, C. L. Degen, and P. Gambardella, Current-Induced Fragmentation of Antiferromagnetic Domains, ArXiv:1912.05287 (2019).

[24] S. Yu. Bodnar, L. Šmejkal, I. Turek, T. Jungwirth, O. Gomonay, J. Sinova, A. A. Sapozhnik, H.-J. Elmers, M. Kläui, and M. Jourdan, Writing and Reading Antiferromagnetic $Mn_2Au$ by Néel Spin-Orbit Torques and Large Anisotropic Magnetoresistance, Nat Commun 9, 348 (2018).

[25] S. Yu. Bodnar, Y. Skourski, O. Gomonay, J. Sinova, M. Kläui, and M. Jourdan, Magnetoresistance Effects in the Metallic Antiferromagnet $Mn_2Au$, Phys. Rev. Applied 14, 014004 (2020).

[26] T. Matalla-Wagner, M.-F. Rath, D. Graulich, J.-M. Schmalhorst, G. Reiss, and M. Meinert, Electrical Néel-Order Switching in Magnetron-Sputtered CuMnAs Thin Films, Phys. Rev. Applied 12, 064003 (2019)

[27] Z. Kašpar, M. Surýnek, J. Zubáč, F. Krizek, V. Novák, R. P. Campion, M. S. Wörnle, P. Gambardella, X. Marti, P. Němec, K. W. Edmonds, S. Reimers, O. J. Amin, F. Maccherozzi, S. S. Dhesi, P. Wadley, J. Wunderlich, K. Olejník, and T. Jungwirth, Quenching of an Antiferromagnet into High Resistivity States Using Electrical or Ultrashort Optical Pulses, Nat Electron 4, 30 (2020).

[28] F. Krizek, Z. Kašpar, A. Vetushka, D. Kriegner, E. M. Fiordaliso, J. Michalicka, O. Man, J. Zubáč, M. Brajer, V. A. Hills, K. W. Edmonds, P. Wadley, R. P. Campion, K. Olejník, T. Jungwirth, and V. Novák, Molecular Beam Epitaxy of CuMnAs, Phys. Rev. Materials 4, 014409 (2020).

[29] Chen HT, O'Hara JF, Azad AK, Taylor AJ. Manipulation of terahertz radiation using metamaterials. Laser Photonics Reviews 5, 513–533 (2011).

[30] M. Sajadi, M. Wolf, and T. Kampfrath, Terahertz-field-induced optical birefringence in common window and substrate materials Optics Express 23, 28985 (2015)

[31] Olejník, K. *et al.* Antiferromagnetic CuMnAs multi-level memory cell with microelectronic compatibility. *Nat Commun* 8, 15434 (2017).

[32] Neu, J., Regan, K. P., Swierk, J. R. & Schmuttenmaer, C. A. Applicability of the thin-film approximation in terahertz photoconductivity measurements. *Appl. Phys. Lett.* 113, 233901 (2018).

[33] L. Nádvorník, M. Borchert, L. Brandt, R. Schlitz, K. A. de Mare, K. Výborný, I. Mertig, G. Jakob, M. Kläui, S. T. B. Goennenwein, M. Wolf, G. Woltersdorf, and T. Kampfrath, Broadband Terahertz Probes of Anisotropic Magnetoresistance Disentangle Extrinsic and Intrinsic Contributions, Phys. Rev. X 11, 021030 (2021).

[34] M. Niemeyer, P. Kleinschmidt, A. W. Walker, L. E. Mundt, C. Timm, R. Lang, T. Hannappel, and D. Lackner, AIP Advances 9, 045034 (2019).

[35] F. Cadiz, D. Lagarde, P. Renucci, D. Paget, T. Amand, H. Carrère, A. C. H. Rowe, and S. Arscott, Appl. Phys. Lett. 110, 082101 (2017).

[36] L. H. Teng, K. Chen, J. H. Wen, W. Z. Lin, and T. S. Lai, J. Phys. D: Appl. Phys. 42, 135111 (2009).

[37] A. Rubano, M. Wolf, and T. Kampfrath, Terahertz conductivity and ultrafast dynamics of photoinduced charge carriers in intrinsic 3C and 6H silicon carbide, Appl. Phys. Lett. 105, 032104 (2014) https://doi.org/10.1063/1.4890619

[38] M. Sajadi, M. Wolf, T. Kampfrath, Terahertz field enhancement via coherent superposition of the pulse sequences after a single optical-rectification crystal, Appl. Phys. Lett. 104, 091118 (2014) https://doi.org/10.1063/1.4867648

[39] E. S. Harmon, M. R. Melloch, J. M. Woodall, D. D. Nolte, N. Otsuka, and C. L. Chang, Carrier Lifetime versus Anneal in Low Temperature Growth GaAs, Appl. Phys. Lett. 63, 2248 (1993).

[40] V. A. Kozlov, F. Y. Soldatenkov, V. G. Danilchenko, V. I. Korolkov, and I. L. Shulpina, Defect Engineering for Carrier Lifetime Control in High Voltage GaAs Power Diodes, in 25th Annual SEMI Advanced Semiconductor Manufacturing Conference (ASMC 2014) (IEEE, Saratoga Springs, NY, 2014), pp. 139–144.

[41] R. I. Stantchev, B. Sun, S. M. Hornett, P. A. Hobson, G. M. Gibson, M. J. Padgett, E. Hendry, Noninvasive, near-field terahertz imaging of hidden objects using a single-pixel detector. Sci. Adv. 2, e1600190 (2016).

[42] D. Weller, G. Parker, O. Mosendz, A. Lyberatos, D. Mitin, N. Y. Safonova, M. Albrecht, Review Article: FePt heat assisted magnetic recording media, J. Vac. Sci. Technol. B 34, 060801 (2016).

[43] W.A. Challener, C. Peng, A. V. Itagi, D. KArns, W. Peng, Y. Peng, X. Yang, X. Zhu, N. J. Gokemeijer, Y.-T. Hsia, G. Ju, R. E. Rottmayer, M. A. Seigler, and E. C. Gage, Heat-assisted magnetic recording by a near-field transducer with efficient optical energy transfer, Nature Photonics 3, 220–224(2009)

[44] C. Zhong, P. Flanigan, N. Abadía, F. Bello, B. D. Jennings, G. Atcheson, J. Li, J.-Y. Zheng, J. J. Wang, R. Hobbs, D. McCloskey, and J. F. Donegan, Effective heat dissipation in an adiabatic near-field transducer for HAMR, Opt. Express 26, 18842-18854 (2018)

[45] M. Tanaka, Recent Progress in Ferromagnetic Semiconductors and Spintronics Devices, Jpn. J. Appl. Phys. 60, 010101 (2021).

[46] S. M. Yakout, Spintronics: Future Technology for New Data Storage and Communication Devices, J Supercond Nov Magn 33, 2557 (2020).

[47] N. A. Spaldin and R. Ramesh, Advances in Magnetoelectric Multiferroics, Nature Mater 18, 203 (2019).